\documentclass[aps,preprint,a4paper,showpacs,showkeys,superscriptaddress]{revtex4-1}
\usepackage{latexsym}
\usepackage{amsmath,amssymb}
\usepackage{graphicx}
\usepackage{subfigure}
\usepackage{xcolor}
\usepackage{cancel}
\usepackage{multirow,tabularx,boldline,array}

\newcolumntype{C}[1]{>{\centering\arraybackslash}p{#1}}
\usepackage{hyperref}     % color hypertext for pdflatex
\hypersetup{colorlinks,%
  citecolor=blue,%
  linkcolor=cyan,%
  pdftex}

\usepackage[titletoc]{appendix}
\usepackage{enumerate}
\begin{document}

%\preprint{arXiv:yymm.nnnn [gr-qc]}

\title{Quantum radiation of a collapsing shell in three-dimensional AdS spacetime revisited}

\author{Hwajin Eom}%
\email[]{um16@sogang.ac.kr}%
\affiliation{Department of Physics, Sogang University, Seoul, 04107,
  Republic of Korea}%
\affiliation{Center for Quantum Spacetime, Sogang University, Seoul 04107, Republic of Korea}%

\author{Wontae Kim}%
\email[]{wtkim@sogang.ac.kr}%
\affiliation{Department of Physics, Sogang University, Seoul, 04107,
	Republic of Korea}%
\affiliation{Center for Quantum Spacetime, Sogang University, Seoul 04107, Republic of Korea}%

\date{\today}

\begin{abstract}
In three-dimensional AdS space, we consider the gravitational collapse of dust shell and then investigate the quantum
radiation from the collapsing shell by employing the functional Schr\"odinger formalism.
In the formation of the BTZ black hole, the interior geometry of the shell can be chosen as
either the massless black hole or the global AdS space.
In the incipient black hole limit, we obtain the wave function exactly
from the time-dependent Schr\"odinger equation for a massless
scalar field. Then, we show that the occupation number of excited states can be
written by analytic expressions, and
the radiation temperature is in agreement with the Hawking temperature,
irrespective of the specific choice of the interior geometries.
\end{abstract}

%

% \pacs{04.70.Dy, 04.62.+v, 04.60.Kz }


\maketitle
%%%%%%%%%%%%%%
%% RevTeX Style End %%
%%%%%%%%%%%%%%

%\newcommand{\lp}{\ell_P}

\raggedbottom

\section{introduction}
\label{sec:introduction}
Since the discovery of Hawking radiation \cite{Hawking:1974sw},
quantum black holes have been one of the most intriguing physical objects in quantum theory of gravity.
In the semiclassical approximations,
the total number of particles emitted from black holes
turned out to be the Planckian distribution of thermal radiation
\cite{Hawking:1974sw,Hartle:1976tp,Gibbons:1977mu,Akhmedov:2015xwa}.

On the other hand, there has been much attention to the collapse of the dust shell for
the Schwarzschild black hole \cite{Boulware:1975fe,Gerlach:1976ji,10.1143/PTP.81.826,Alberghi:1998xe,Alberghi:2001cm}
and lower-dimensional black holes \cite{Mann:1992my,Ross:1992ba,Peleg:1994wx,Crisostomo:2003xz,Mann:2006yu,Hyun:2006xt}.
In particular, the functional Schr\"odinger formalism  was employed
to investigate quantum radiation from the collapsing shell for the Schwarzschild metric
\cite{Vachaspati:2006ki,Vachaspati:2007hr,Greenwood:2008zg,Greenwood:2010sx,Kolopanis:2013sty,Saini:2015dea,Saini:2016rrt}.
In fact,  the advantage of the functional Schr\"odinger
equation would be
that wave functionals
living on a space of field configurations explicitly depend on time \cite{Kiefer:1991xy}.
This functional Schr\"odinger formalism
was also applied to the formation of various dynamical geometries such as
the Reissner-Nordstr\"om black hole \cite{Greenwood:2009pd,Das:2019iru},
the black string in anti-de Sitter(AdS) space \cite{Greenwood:2009gp},
the three-dimensional AdS black hole \cite{Saini:2017mur},
and the regular black hole \cite{Um:2019qgc}.
In the formation of the black holes,
the spectrum of quantum radiation is approximately thermal
as the shell approaches its own horizon,
and the temperature of the radiation is in agreement with
the Hawking temperature of the corresponding black holes.
%BTZ black string \cite{Greenwood:2009gp}
%numerically
%\cite{Vachaspati:2006ki,Vachaspati:2007hr,Greenwood:2008zg,Greenwood:2009pd,Greenwood:2010sx,Saini:2016rrt,Saini:2017mur}
%analytically
%\cite{Kolopanis:2013sty,Saini:2015dea,Um:2019qgc}.
In asymptotically flat black holes formed by collapsing shells,
spacetime for the interior region of the shell
was naturally prescribed by Minkowski spacetime because the interior region is assumed to be empty \cite{Vachaspati:2006ki,Vachaspati:2007hr,Greenwood:2008zg,Greenwood:2009pd,Greenwood:2010sx,Kolopanis:2013sty,Saini:2015dea,Saini:2016rrt,Das:2019iru,Um:2019qgc}.

In the asymptotically non-flat spacetime such as
the nonrotating
Ba\~nados-Teitelboim-Zanelli (BTZ) black hole \cite{Banados:1992wn}
described by the line element
\begin{equation}
\label{eq:metric of BTZ bh}
ds^2=-f(r) dt^2 +\frac{1}{f(r)}dr^2 +r^2 d\phi^2,
\end{equation}
where $f(r) =-M +r^2/\ell^2$
with $M$ and $\ell$ being the mass parameter and the radius of curvature,
respectively, the BTZ black hole is defined for $M>0$ and the massless black hole is for $M=0$.
In addition, the global AdS space of a thermal soliton emerges as a bound state
for $M=-1$ which has a lower energy than that of the massless black hole.
In the formation of the BTZ black hole, the interior region of the shell could be chosen
as the global AdS space ($M=-1$) \cite{Peleg:1994wx,Crisostomo:2003xz} or the massless black hole geometry ($M=0$)
\cite{Saini:2017mur}. For $M=0$,
the radiation temperature was shown to be close to the Hawking temperature by explicit
numerical analysis \cite{Saini:2017mur}.

In this paper,
we will revisit the quantum radiation by solving the functional Schr\"odinger equation
analytically in the formation of the BTZ black hole.
For this purpose,
the exterior and interior regions of the dust shell will be described by the AdS geometry \eqref{eq:metric of BTZ bh}
characterized by different exterior and interior mass parameters of $M_{\rm out}$ and $M_{\rm in}$.
The mass parameters are chosen as
$M_{\rm out} >0$ and $M_{\rm in} =0,-1$.
Namely, the vacuum state for the interior region
is described by  the massless black hole or the global AdS space.
Importantly, in the incipient limit of $R\to R_H$
where $R$ is a shell radius, the occupation number will be expressed by
explicit Bessel functions
and it turns out to be independent of $M_{\rm in}$.
After all, the radiation temperature is found to be close to the Hawking temperature.

The organization of this paper is as follows.
In Sec.~\ref{sec:general},
we will find a mass relation from the junction condition on the dust shell
for arbitrary mass parameters of $M_{\rm out}$ and $M_{\rm in}$.
%which correspond to the exterior and the interior regions of the shell, respectively.
Then, the functional Schr\"odinger equation from the action for a massless scalar field
will be derived for the formation of the BTZ black hole.
In Sec.~\ref{sec:cases},
the wave functions from the functional Schr\"odinger equations
will be obtained analytically and the corresponding spectra of quantum radiation will be investigated.
Finally, conclusion and discussion will be given in Sec.~\ref{sec:conclusion}.

\section{Schr\"odinger equations for collapsing dust shell}
\label{sec:general}
\subsection{Classical collapse of the shell}
\label{sec:general classical}
%3 dimensional anti de Sitter space, the cosmological constant is $\Lambda=-\ell^{-2}$

%analogous to infinitely thin shell \cite{Vachaspati:2006ki,Vachaspati:2007hr,Greenwood:2008zg,Greenwood:2009gp,Greenwood:2009pd,Greenwood:2010sx,Saini:2015dea,Saini:2016rrt,Saini:2017mur,Das:2019iru,Um:2019qgc}
In this section, we provide a general formalism for the classical collapse of the shell in AdS space.
Let us start with  spacetime divided by an infinitely thin shell with a radius $R$ in three-dimensional AdS space.
For the exterior and interior coordinates denoted by $\left(t,r,\phi\right)$
and $\left(T,r,\phi\right)$,
%respectively,
the radius would be given as
$R(t)$ which is a decreasing function of the time $t$.

For the exterior region of the shell, the spacetime is given by the line element
    \begin{equation}
    \label{eq:exterior}
    ds_{\rm (out)}^2 = -f_{\rm out}(r) dt^2+\frac{1}{f_{\rm out}(r)} dr^2 +r^2 d\phi^2~~(r>R(t)),
    \end{equation}
where $f_{\rm out}(r) = -M_{\rm out}+r^2/\ell^2$.
For the interior region of the shell, the line element is also described by
	\begin{equation}
    \label{eq:interior}
	ds_{\rm (in)}^2  = -f_{\rm in}(r) dT^2 + \frac{1}{f_{\rm in}(r)} dr^2 +r^2 d\phi^2~~(r<R(t)),
	\end{equation}
where $f_{\rm in} (r) =-M_{\rm in}+r^2/\ell^2$.
Thus, the line element at the shell is assumed to be
    \begin{equation}
    \label{eq:shell}
    ds_{\rm (shell)}^2 =-d\tau^2+ R^2(t) d\phi^2~~(r=R(t))
    \end{equation}
with $\tau$ being the proper time of the collapsing shell.
Since the line element should be continuous at the shell,
Eqs.~\eqref{eq:exterior} and \eqref{eq:interior} are coincident with Eq.~\eqref{eq:shell} at $r=R(t)$,
which gives the relations between the time coordinates $t$ and $T$ as
    \begin{equation}
    \label{eq:time rel}
    \lambda(t)=\frac{dT}{dt}  \bigg|_{r=R(t)}= \sqrt{ \frac{f_{\rm out} (R)}{f_{\rm in} (R)}-
    \frac{1}{f_{\rm in}(R)} \left\{ \frac{1}{f_{\rm out}(R)} -\frac{1}{f_{\rm in}(R)} \right\} \left(\frac{dR}{dt}\right)^2
    },
    \end{equation}
where we did not fix $f_{\rm out}(R)$ and $f_{\rm in}(R)$ for our general argument.

Next, we consider a combined metric tensor defined as
$g_{\mu\nu} =\Theta(r-R(t)) g^{\rm (out)}_{\mu\nu} + \Theta(-r+R(t)) g^{\rm (in)}_{\mu\nu}$
where $\Theta(x)=1 (x>0)$ and $\Theta(x)=0 (x<0)$.
Employing the first junction condition that the combined metric tensor is
continuous on the shell,
one can derive the projection of the Einstein equation onto the shell as \cite{Israel:1966rt}
    \begin{equation}
    \label{eq:Einstein eq}
    -\left[K_{\mu\nu}\right]+h_{\mu\nu} \left[K\right] = 8\pi S_{\mu\nu},
    \end{equation}
where $K_{\mu\nu}=h_\mu^\gamma \nabla_\gamma n_\nu$, $K=g_\mu^\gamma K_{\gamma\nu}$, and $S_{\mu\nu}=h_\mu^\gamma T_{\gamma \nu}$
are the projections of the extrinsic curvature and the energy-momentum tensor of the shell, respectively.
$[...]$ denotes
$[A] = \lim_{r\to R+} A(r) - \lim_{r\to R-} A(r)$.
And $h_{\mu\nu}=(g_{\mu\nu} -n_\mu n_\nu) \big|_{r=R(t)}$
is a projection operator onto the shell with a normal vector of the shell $n_\mu$,
where $h_{\mu\nu}$ automatically satisfies $h_\mu^\gamma h_{\gamma\nu} = h_{\mu\nu}$.
We consider the dust shell and so its energy-momentum tensor is given as $S^{\mu\nu}=\sigma u^\mu u^\nu$, where $\sigma$ is the surface energy density of
the shell and $u^\mu$ is the three-velocity at the shell.
Taking the inner product to Eq.~\eqref{eq:Einstein eq} with $u^\mu$ and $u^\nu$,
one can finally get the so-called mass relation,
    \begin{equation}
    \label{eq:mass rel}
    \sqrt{\dot R^2 + f_{\rm out}(R)}-\sqrt{\dot R^2 + f_{\rm in}(R)} =-8\pi\sigma R,
    \end{equation}
where $\dot R=dR/d\tau$. Note that Eq.~\eqref{eq:mass rel}
certainly shows that
$\sigma>0$ for $M_{\rm out} > M_{\rm in}$
while $\sigma<0$ for $M_{\rm out} < M_{\rm in}$.

Plugging $f_{\rm out}(r)$ and $f_{\rm in}(r)$ in Eqs.~\eqref{eq:exterior} and \eqref{eq:interior} into Eq.~\eqref{eq:mass rel}, one can find
    \begin{equation}
    \label{eq:R dot square}
    \dot R^2 =\frac 1 2 \left(M_{\rm out}+ M_{\rm in} \right)
    +\left(\frac{f_{\rm out}(R)- f_{\rm in} (R)}{16\pi\sigma R} \right)^2
    +\left(\left(4\pi\sigma\right)^2-\frac{1}{\ell^2} \right) R^2,
    \end{equation}
where we supposed that $(4\pi\sigma)^2 > \ell^{-2}$ to make the shell collapse continuously even
for an arbitrary large $R$.
Using the relation of $\left(dR/dt\right)^2 = \left(dt/d\tau\right)^{-2} \dot R^2$
for the collapsing shell, we get
    \begin{equation}
    \label{eq:dR over dt}
    \frac{dR}{dt}=-\frac{f_{\rm out}(R) \dot R}{\sqrt{f_{\rm out}(R)+\dot R^2}}.
    \end{equation}

\subsection{The functional Schr\"odinger equation}
\label{sec:general quantum}
To study the quantum radiation
from the collapsing shell, we consider a minimally coupled scalar field $\Phi$.
On the spacetime background described by Eqs.~\eqref{eq:exterior}, \eqref{eq:interior}, and \eqref{eq:shell},
we consider the total action which consists of the sum of the exterior and interior actions as
    \begin{align}
    \label{eq:total action}
    & S_{\rm (tot)} = S_{\rm (out)} +S_{\rm (in)}, \\ \label{eq:exterior action}
    & S_{\rm (out)} =\pi \int dt \int_{R(t)}^\infty  \!\!\!\!dr~\left(\frac{1}{f_{\rm out}(r)} \left(\partial_t \Phi\right)^2 -f_{\rm out}(r)\left(\partial_r \Phi\right)^2 \right),\\ \label{eq:interior action}
    & S_{\rm (in)} =\pi \int dT \int^{R(t)}_0 \!\!\!\!\!\!dr~ \left(\frac{1}{f_{\rm in}(r)} \left(\partial_T \Phi\right)^2 -f_{\rm in}(r)\left(\partial_r \Phi\right)^2 \right).
    \end{align}
For convenience, let us rewrite the interior action by using Eq.~\eqref{eq:time rel} as
    \begin{equation}
    \label{eq:interior action}
    S_{\rm (in)} =\pi \int dt \int^{R(t)}_0 \!\!\!\!\!\!dr~\left(\frac{1}{\lambda(t) f_{\rm in}(r)} \left(\partial_t \Phi\right)^2 - \lambda(t) f_{\rm in}(r)\left(\partial_r \Phi\right)^2 \right).
    \end{equation}
From Eq.~\eqref{eq:total action}, the functional Schr\"odinger equation will be derived
in the limit of $\lambda(t) \ll 1$,
which will be shown to be of relevance to the incipient limit of the BTZ black hole of $R\to R_H$.

%\subsubsection{Formation of the BTZ black hole : $M_{\rm out}>0$, $M_{\rm in}=0,-1$}
In order to study the formation of the BTZ black hole, we
consider the configuration of $M_{\rm out}> 0$ and $M_{\rm in}=0,-1$.
In particular, for $M=-1$,
nonsingular vortex solutions in AdS$_3$ background were numerically found \cite{Edery:2020kof}.
However, we will consider the global AdS space as a thermal soliton
without any classical fields.
As the radius of the shell approaches its horizon, $i.e.$, $R(t) \to R_H =\sqrt{M_{\rm out}}\ell$ called the incipient limit,
Eq.~\eqref{eq:R dot square} reduces to a finite value as
$\dot R^2 \approx \left.\left(f_{\rm in}(R)/(8\mu)-2\mu\right)^2\right|_{R=R_H}$,
where $\mu$ is the constant of motion defined by $\mu =2 \pi R \sigma$ \cite{Saini:2017mur}.
Thus, Eq.~\eqref{eq:dR over dt} in this limit also reduces to a simpler form as
    \begin{equation}
    \label{eq:dR over dt sim}
    \frac{dR}{dt} \approx -f_{\rm out} (R),
    \end{equation}
which yields the radius function as
    \begin{equation}
    \label{eq:radius function}
    R(t) =
        \begin{cases}
        R_0 & (t<0), \\
        R_H +\left(R_0-R_H \right) e^{-f_{\rm out}'(R_H) t} & ( 0\le t\le t_f),\\
        R_f & (t>t_f)~,
        \end{cases}
    \end{equation}
%$R(t)=R_H +\left(R_0-R_H \right) e^{-f_{\rm out}'(R_H) t}$ for $0\le t\le t_f$
where $t=0$ and $R_0$ are the initial time and radius, respectively.
For a nice mathematical manipulation of infinite time,
we artificially take the collapse to stop at some time
$t=t_f$ with the radius $R_f$
defined as $R(t_f)=R_H +(R_0 -R_H) e^{-f'_{\rm out} (R_H) t_f}$,
where Eq.~\eqref{eq:radius function} is continuous at $t=t_f$, $i.e.$, $R_f=R(t_f)$.
Then we take the limit of $t_f \to\infty$ to investigate the eternal collapse case
which implies that the shell collapses to form the black hole ($i.e.$, $R(t)\to R_H$),
in other words, $R_f \to R_H$ in Eq.~\eqref{eq:radius function}.
Note that,  if one considered up to the order of $(R-R_H)^2$ in Eq.~\eqref{eq:mass rel},
$M_{\rm in}$ would appear in $(dR/dt)$ in such a way that
$(dR/dt)\approx -f_{\rm out}(R)+f_{\rm out}^2(R)/(2 \dot R^2)$.
However, as long as we are concerned with the lowest order of $(R-R_H)$,
$i.e.$, in the incipient limit, $(dR/dt)$ turns out to be independent of $M_{\rm in}$.
Therefore, in the formation of the BTZ black hole,
the choice of the interior region is not crucial as long as $M_{\rm out} > M_{\rm in}$.

From Eq.~\eqref{eq:dR over dt},
Eq.~\eqref{eq:time rel} can be cast in the incipient limit as
    \begin{equation}
    %\label{eq:lambda w epsilon}
    \label{eq:approx lambda}
    \lambda(t)
    %& \!=\!\!\left[
    %\frac{f_{\rm out}(R)}{f_{\rm in} (R)}
    %-\frac{f_{\rm out}(R)}{f_{\rm in} (R)} \left(
    %1-\frac{f_{\rm out}(R)}{f_{\rm in} (R)} \right)
    %\left\{1
    %-\left(\frac{f_{\rm in}(R_H)}{8\mu}-2\mu \right)^{-2} \!\!f_{\rm out}(R)  \right\}\right]^{\frac 1 2}\\
    %&
    \approx \frac{16\mu^2+f_{\rm in}(R_H)}{\left(16\mu^2-f_{\rm in}(R_H)\right)f_{\rm in}(R_H)} f_{\rm out}(R(t))
    \end{equation}
so that $\lambda(t)$ can be made very small
because $f_{\rm out}\to 0$ near the horizon.
For $\lambda(t)\ll 1$, the kinetic term in the exterior action \eqref{eq:exterior action} and the potential term in the interior action \eqref{eq:interior action}
can be ignored, and thus, the total action \eqref{eq:total action} is simplified as
    \begin{equation}
    \label{eq:tot action sim}
    S_{\rm (tot)}=\pi \int dt \left(\frac{1}{\lambda(t)} \int_0^{R(t)} \!\!\!\!\!\!dr~ \frac{1}{f_{\rm in} (r)} \left(\partial_t \Phi\right)^2-
    \int_{R(t)}^\infty \!\!\!\!dr~ f_{\rm out}(r) \left(\partial_r \Phi\right)^2 \right).
    \end{equation}
After decomposition of $\Phi$ in terms of $\Phi =\sum_{k} a_k (t) d_k (r)$
and the simultaneous diagonalization of the following Hermitian operators \cite{goldstein:mechanics,Vachaspati:2006ki},
$A_{k,k'} =2\pi\int_0^{R(t)} dr~ d_k(r) d_{k'}(r)/f_{\rm in} (r)$ and
$B_{k,k'} =2\pi\int_{R(t)}^\infty dr~f_{\rm out}(r)(dd_k (r)/dr)(dd_{k'} (r)/dr)$,
we obtain the total action \eqref{eq:tot action sim} as
    \begin{eqnarray}
    \notag
    S_{\rm (tot)} & =&\int dt L_\Phi \\
    \label{eq:tot action diag}
    &=& \sum_k \int dt \left(\frac{\alpha_k}{2\lambda(t)} \left(\frac{d b_k(t)}{dt}\right)^2 -\frac 1 2 \beta_k b_k^2(t) \right),
    \end{eqnarray}
where $\alpha_k$ and $\beta_k$ are eigenvalues of $A_{k,k'}$ and $B_{k,k'}$
for a set of eigenvectors $\{b_k\}$, respectively.
For a mode amplitude $b_k(t)$,
the conjugate momentum is defined as $\Pi_k=\partial L_\Phi/\partial \left(d b_k(t)/dt\right)$
$=\alpha_k \lambda^{-1}(t) \left(db_k (t)/dt\right)$.
Then, from the Legendre transformation of Eq.~\eqref{eq:tot action diag}, the Hamiltonian can easily be obtained
as $H_\Phi=\sum_k \left\{\lambda(t)/\left(2\alpha_k\right) \Pi_k^2+(1/2) \beta_k b_k^2\right\}$.
Imposing the quantization rule $\left[b_k(t),\Pi_{k'}\right]=i\delta_{k,k'}$
where $\Pi_k \to -i\partial/\partial b_k$, we get the Hamiltonian for quantum radiation as
    \begin{equation}
    H_\Phi = \sum_k H_{\Phi,k}
    =\sum_k \left\{-\frac{\lambda(t)}{2\alpha_k} \frac{\partial^2}{\partial b_k^2}+\frac 1 2 \beta_k b_k^2\right\}.
    \end{equation}
It is worthy to note that the Hamiltonian is separable for each mode $k$
so that the functional Schr\"odinger equation of $H_\Phi \Psi= i \partial \Psi /\partial t$ \cite{Vachaspati:2006ki,Vachaspati:2007hr,Greenwood:2008zg,Greenwood:2009gp,Greenwood:2009pd,Greenwood:2010sx,Kolopanis:2013sty,Saini:2015dea,Saini:2016rrt,Saini:2017mur,Das:2019iru,Um:2019qgc}
can be rewritten as the family of the equation of $H_{\Phi,k} \psi_k= i \partial \psi_k /\partial t$
for a certain mode $k$,
where $\Psi = \prod_k \psi_k (t,b_k)$.
Therefore, for one eigenvector $b \in\{b_k\}$ for which the wave function $\psi (t,b) \in \{\psi_k(t,b_k)\}$,
the functional Schr\"odinger equation is written as \cite{Vachaspati:2006ki}
    \begin{equation}
    \label{eq:Sch eq eta}
    \left\{-\frac{1}{2\alpha} \frac{\partial^2}{\partial b^2}
    +\frac 1 2 \alpha \frac{\omega^2_0}{\lambda(\eta(t))} b^2 \right\}\psi(\eta,b)= i\frac{\partial \psi}{\partial \eta} (\eta, b)
    \end{equation}
%with $\omega(\eta)=\omega_0 /\sqrt{\lambda({\displaystyle\eta(t)})}$
in terms of a time parameter $\eta = \int_0^t d t'~ \lambda(t')$,
where $\omega_0 = \sqrt{\beta / \alpha }$
with $\alpha\in\{\alpha_k \}$ and $\beta\in\{\beta_k \}$.
Using Eqs.~\eqref{eq:dR over dt sim} and \eqref{eq:approx lambda},
$R(\eta)$ can also be written in the form of
    \begin{equation}
    \label{eq:Radius eta}
    R(\eta) =\left\{
    \begin{matrix}
        R_0 & (\eta<0), \\
        R_0 -\dfrac{\left(16\mu^2-f_{\rm in}(R_H)\right)f_{\rm in}(R_H)}{16\mu^2+f_{\rm in}(R_H)}\eta & ( 0\le\eta\le \eta_f),\\
        R_f & (\eta> \eta_f),~
    \end{matrix}
    \right.
    \end{equation}
which will be used in Sec.~\ref{sec:cases}
in order to solve the functional Schr\"odinger equation \eqref{eq:Sch eq eta} analytically.

\section{Quantum radiation from the collapsing shell}
\label{sec:cases}
%\subsection{Formation of the BTZ black hole : $M_{\rm out}> 0$, $M_{\rm in}=0,-1$}
%\label{sec:BH}
\subsection{Exact solution of the functional Schr\"odinger equation}
\label{sec:exact sol BH}

The functional Schr\"odinger equation \eqref{eq:Sch eq eta}
can be used to study the quantum radiation of the collapsing BTZ black hole.

For $\eta<0$,
the radius \eqref{eq:Radius eta} is constant, and thus, the wave function $\psi(\eta,b)$ can be given by
the linear combination of the eigenstates of the simple harmonic oscillator as
    \begin{equation}
    \label{eq:eigenstate SHO}
    \varphi_n (b) =\left( \frac{\alpha \bar \omega}{\pi}\right)^{\frac 1 4} \frac{1}{\sqrt{2^n n!}} H_n (\sqrt{\alpha \bar \omega} b) e^{-\frac 1 2 \alpha \bar \omega b^2},
    \end{equation}
where $H_n (x)$ is a Hermite polynomial and
$\bar \omega=\omega_0$.
Starting from a vacuum state represented by a quantum state with the lowest energy,
we set $\psi(\eta,b)=e^{-i \omega_0 \eta} \varphi_0(b)\big|_{\bar \omega =\omega_0}$.

For $0\leq\eta \leq\eta_f$,
let us obtain the exact solution to satisfy Eq.~\eqref{eq:Sch eq eta}.
By plugging $R(\eta)$ in Eq.~\eqref{eq:Radius eta} into Eq.~\eqref{eq:approx lambda}, $\lambda(t)$ in Eq.~\eqref{eq:approx lambda} can be rewritten as
    \begin{equation}
    \label{eq:lambda eta}
    \lambda(\eta)=\frac{d\eta}{dt}
    =\frac{f'_{\rm out}(R_H)}{H} \left(1-H\eta\right)
    \end{equation}
where
$H=\{\left(16\mu^2+f_{\rm in}(R_H)\right)(R_0-R_H)\} / \{\left(16\mu^2-f_{\rm in}(R_H)\right)f_{\rm in}(R_H)\}$.
Using Eq.~\eqref{eq:lambda eta} with the following redefined relations of
$\hat \alpha = H\alpha$, $\hat \omega_0 = \omega_0 /\sqrt{f'_{\rm out}(R_H) H}$,
and $\hat \eta=H \eta$,
one can simplify the functional Schr\"odinger equation in the incipient limit as \cite{Kolopanis:2013sty}
    \begin{equation}
    \label{eq:Sch eq eta hat}
    \left\{-\frac{1}{2\hat \alpha} \frac{\partial^2}{\partial b^2} +
    \frac 1 2 \hat \alpha \frac{\hat \omega_0^2}{1-\hat \eta}\right\} \psi=
    i\frac{\partial\psi}{\partial \hat\eta}.
    \end{equation}
The exact solution to Eq.~\eqref{eq:Sch eq eta hat}
 is known as \cite{Dantas:1990rk}
    \begin{equation}
    \label{eq:solution}
    \psi(\hat \eta,b)=e^{i \theta(\hat \eta)} \left(\frac{\hat \alpha}{\pi \zeta^2}\right)^{\frac 1 4} e^{i\left(\frac{1}{\zeta} \zeta_{\hat \eta}+\frac{i}{\zeta^2}\right) \frac{\alpha b^2}{2}},
    \end{equation}
where $\theta(\hat \eta)=(-1/2)\int_0^{\hat \eta} d \hat \eta ' \zeta^{-2} (\hat \eta ')$.
Note that $\zeta(\hat \eta)$ is the solution to the inhomogeneous nonlinear equation of
    \begin{equation}
    \label{eq:soluble diff eq}
    \frac{d ^2 \zeta}{d \hat \eta^2}+\frac{\hat \omega_0 ^2}{1-\hat \eta} \zeta=\frac{1}{\zeta^3}
    \end{equation}
with the initial conditions: $\zeta(0)=1/\sqrt{\hat \omega_0}$ and
$\zeta_{\hat \eta}(0)=0$,
where $\zeta_{\hat \eta}=d\zeta/d\hat \eta$.
In the previous work for the formation of the BTZ black hole with $M_{\rm in}=0$
\cite{Saini:2017mur},
Eq.~\eqref{eq:soluble diff eq} was solved numerically.
Instead, we will find the analytic expression for Eq.~\eqref{eq:soluble diff eq}.

Let us now suppose that the solution satisfying Eq.~\eqref{eq:soluble diff eq} is written as \cite{Kolopanis:2013sty}
    \begin{equation}
    \label{eq:solution1}
    \zeta(\hat \eta)=\frac{1}{\sqrt{\hat \omega_0}} \sqrt{\xi^2 (\hat \eta)+\chi^2 (\hat \eta) }
    \end{equation}
with the appropriate initial conditions:
$\xi(0)=1,~\xi_{\hat \eta}(0)=0,~\chi(0)=0,~\chi_{\hat \eta} (0)=\hat \omega_0$,
and its derivative with respect to $\hat \eta$ is also written as
$\zeta_{\hat \eta} =\left(\xi \xi_{\hat \eta}+\chi \chi_{\hat \eta}\right)/(\hat \omega_0 \zeta)$.
In Eq.~\eqref{eq:soluble diff eq}, the two linearly independent homogeneous solutions
denoted by $y_1(\hat \omega_0,\hat \eta)$ and $y_2(\hat \omega_0,\hat \eta)$ are obtained as
the Bessel function of the first kind $J_1 (2\hat \omega_0 \sqrt{1-\hat \eta})$ and
the Bessel function of the second kind $Y_1 (2\hat \omega_0 \sqrt{1-\hat \eta})$, respectively \cite{10.5555/1098650}.
Then, we can construct $\xi(\hat \eta)$ and $\chi(\hat \eta)$ in Eq.~\eqref{eq:solution1} as
    \begin{eqnarray}
    \label{eq:wrons1}
    \xi(\hat \omega_0,\hat \eta) =& W^{-1}
    \left( y_{2\hat \eta}(\hat \omega_0,0) y_1 (\hat \omega_0,\hat \eta)\right.
    \left. -y_{1\hat \eta}(\hat \omega_0,0) y_2 (\hat \omega_0,\hat \eta)\right),\\
    \chi(\hat \omega_0,\hat \eta) =& -\hat \omega_0 W^{-1}
    \left( y_{2}(\hat \omega_0,0) y_1 (\hat \omega_0,\hat \eta)\right.
    \left. -y_{1}(\hat \omega_0,0) y_2 (\hat \omega_0,\hat \eta)\right),\\
    \xi_{\hat \eta} (\hat \omega_0,\hat \eta) = &W^{-1}
    \left( y_{2\hat \eta}(\hat \omega_0,0) y_{1\hat \eta} (\hat \omega_0,\hat \eta)\right.
    \left. -y_{1\hat \eta}(\hat \omega_0,0) y_{2 \hat \eta} (\hat \omega_0,\hat \eta)\right),\\
    \label{eq:wrons4}
    \chi_{\hat \eta} (\hat \omega_0,\hat \eta) = &-\hat \omega_0 W^{-1}
    \left( y_2(\hat \omega_0,0) y_{1\hat \eta} (\hat \omega_0,\hat \eta)\right.
    \left. -y_1(\hat \omega_0,0) y_{2 \hat \eta} (\hat \omega_0,\hat \eta)\right),
    \end{eqnarray}
where $W$ is the Wronskian at the initial time calculated as $W=y_{2\hat \eta}(\hat\omega_0,0) y_1(\hat\omega_0,0)-
y_{1\hat \eta}(\hat\omega_0,0) y_2(\hat\omega_0,0)=-\pi^{-1}$.
From Eqs.~\eqref{eq:wrons1}-\eqref{eq:wrons4},
we explicitly get %$\xi(\hat \eta)$ and $\chi(\hat \eta)$ are
    \begin{eqnarray}
    \label{eq:sol1}
    \xi (x,z) =& \phantom{-}\dfrac{\pi z}{2}&\left(Y_0 (x) J_1(z)-J_0(x) Y_1(z) \right),\\ \label{eq:sol2}
    \chi (x,z) =& \phantom{-}\dfrac{\pi z}{2}&\left(Y_1 (x) J_1(z)-J_1(x) Y_1(z) \right),\\ \label{eq:sol3}
    \xi_{\hat \eta} (x,z) =& -\dfrac{ \pi x^2}{4} &\left(Y_0 (x) J_0(z) -J_0 (x) Y_0 (z)\right),\\ \label{eq:sol4}
    \chi_{\hat \eta} (x,z)  =& -\dfrac{ \pi x^2}{4} &\left(Y_1 (x) J_0(z) -J_1 (x) Y_0 (z)\right),
    \end{eqnarray}
where the new variables are defined by
    \begin{equation}
    \label{eq:x and z BH}
    x=2\hat \omega_0,\qquad z=2\hat \omega_0 \sqrt{1-\hat \eta}.
    \end{equation}
Consequently, we obtain the analytic solution $\zeta(\hat \eta)$ and its derivative $\zeta_{\hat \eta}$,
    \begin{equation}
    \label{eq:sol tot}
    \zeta(x,z)=\sqrt{\frac 2 x \big(\xi^2 +\chi^2\big)},\quad
    \zeta_{\hat \eta}(x,z)=\sqrt{\frac 2 x}~ \frac{\xi \xi_{\hat \eta}+\chi \chi_{\hat \eta}}{\sqrt{\xi^2 +\chi^2 }}
        \end{equation}
in terms of the combinations of the Bessel functions.

Finally, if $\eta>\eta_f$, then $R(\eta)= R_f$,
and thus, Eq.~\eqref{eq:Sch eq eta} becomes the Schr\"odinger equation of the time-independent simple harmonic oscillator
so that the wave function is simply given as
$\psi(\eta,b)=\left(\alpha \omega_f /\pi\right)^{\frac 1 4} e^{-i \omega_f \eta_f}  e^{-\frac 1 2 \alpha  \omega_f b^2}$,
where $\omega_f=\omega_0/\sqrt{\lambda(\eta_f)}$.

\subsection{Spectrum of the quantum radiation}
\label{sec:spectrum BH}
In order to study the quantum radiation from the shell,
we should calculate the occupation number defined by
    \begin{equation}
    \label{eq:occ num def}
    N=\sum_{n=0}^\infty n |c_n (\eta) |^2,
    \end{equation}
where the probability amplitude in the $n$th eigenstate \eqref{eq:eigenstate SHO} for the frequency
$\bar \omega =\omega_f$
can be calculated as
    \begin{align}
    c_{n} (\eta)
    &= \int_{-\infty}^{\infty} \textrm d b~ \varphi_n (b)  \psi(\eta,b)\\
    &=\left\{
        \begin{array}{ll}
        e^{i \theta(\hat \eta)} \frac{(n-1)!!}{\sqrt{n!}} \frac{1}{\sqrt[4]{\hat \omega_f \zeta^2 } }
    \sqrt{(-1)^n \frac 2 P \left(1-\frac 2 P\right)^n}  & \qquad (n=0,2,4,...), \\
        0 & \qquad (n=1,3,5,...),\label{eq:prob amp}
        \end{array}
    \right.
    \end{align}
where $P=1-i(\zeta_{\hat \eta}/\hat \omega_f \zeta) + 1/(\hat \omega_f \zeta^2)$ with $\hat \omega_f = \hat \omega_0 /\sqrt{1-H \eta_f}$.
After some calculations, we obtain
    \begin{equation}
    \label{eq:occ num formal}
    N(\hat \eta_f,\hat \omega_0)=\frac{\hat \omega_f \zeta^2}{4}
    \left\{\left(1-\frac{1}{\hat \omega_f \zeta^2}\right)^2
    +\left(\frac{\zeta_{\hat \eta}}{\hat \omega_f \zeta}\right)^2\right\}.
    \end{equation}
%given as $N(\eta_f, \omega_f)$, an explicit function of $\eta_f(=\hat \eta_f /H)$ and %$\omega_f(=H\hat \omega_f)$,
Using the exact
expression \eqref{eq:sol tot},
we can rewrite Eq.~\eqref{eq:occ num formal} as
    \begin{equation}
    \label{eq:occ num BH finite}
    N(x,z) = \left\{\frac x z (\xi^2 +\chi^2)\right\}^{-1}
    \left[~\frac 1 4 \!\left(\frac{x}{z} (\xi^2 +\chi^2)-1\right)^2
    +\left(\frac{\xi\xi_{\hat \eta}+\chi\chi_{\hat \eta}}{x}\right)^2\right].
    \end{equation}
%where we used Eqs.~\eqref{eq:x and z BH} and \eqref{eq:sol tot}.
By integrating Eq.~\eqref{eq:lambda eta},
the relation between $t$ and $\eta$ in the incipient limit can easily be obtained as
$1-H\eta = e^{-f'_{\rm out}(R_H)\phantom{`}t}$
and $\Omega_f$ is expressed by
$\Omega_f = \lambda(t_f) \omega_f = (f'_{\rm out}(R_H)/H) e^{-f'_{\rm out} (R_H)}\omega_f$.
In this limit, Eq.~\eqref{eq:x and z BH} can also be rewritten in terms of $t_f$ and $\Omega_f$ as
    \begin{equation}
    \label{eq:x and z}
    x=\frac{2\Omega_f}{f'_{\rm out} (R_H)} e^{\frac 1 2 f'_{\rm out}(R_H) t_f},\quad
    z=\frac{2\Omega_f}{f'_{\rm out} (R_H)}.
    \end{equation}
Thus, the occupation number becomes the functions of $t_f$ and
$\Omega_f$, {\it i.e.}, $N(x,z)=N(t_f,\Omega_f)$.
\begin{figure}[b]
\centering
\includegraphics[width=0.6\textwidth]{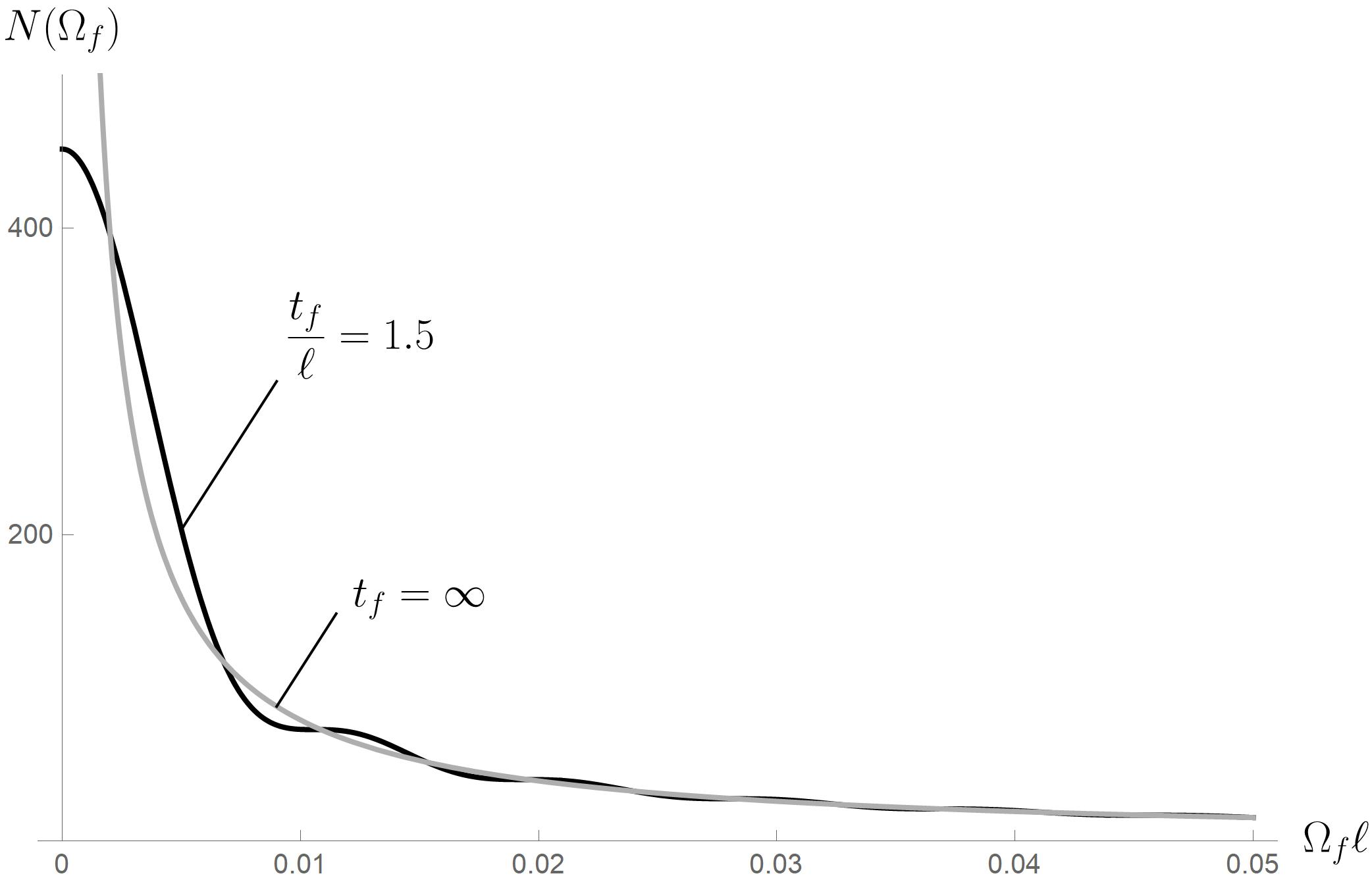}
\caption{The spectrum of the occupation number is plotted
with respect to the dimensionless frequency $\Omega_f \ell$
for two different times.
The black and grey curves are for a finite time \eqref{eq:occ num BH finite} ($t_f/\ell=1.5$) and
for infinite time \eqref{eq:occ num BH infinite}, respectively.
Although the spectrum for the finite time is damped oscillating, the spectrum turns out to be
approximately thermal
for $t_f \to \infty$ when the shell approaches the horizon.
}\label{fig:black hole}
\end{figure}

Note that
the functional Schr\"odinger equation \eqref{eq:Sch eq eta}
tells us that
the wave function is apparently responsible for the exterior and interior geometries through Eq.~\eqref{eq:approx lambda};
however, the occupation number $N(t_f,\Omega_f)$ in the incipient limit is of no relevance to the interior geometry
because $x$ and $z$ are independent of $f_{\rm in}$ as seen from Eq.~\eqref{eq:x and z}.
For the asymptotic observer, the shell does not collapse to $R_H$ in a finite time, but the collapsing shell takes infinite time to reach $R_H$.
So, we should take the limit of $t_f \to \infty$ to study the quantum radiation
when the black hole forms.
For $t_f \to\infty$, we finally obtain the occupation number \eqref{eq:occ num BH finite} as
    \begin{equation}
    \label{eq:occ num BH infinite}
    N_{\infty} (z)= \frac \pi 8 \frac{z}{J_1^2(z)+Y_1^2(z)}
    \left[
    \left(J_1^2(z)+Y_1^2(z)-\frac{2}{\pi z}
    \right)^2
    +\left(J_0(z) J_1(z)+Y_0(z) Y_1(z)\right)^2
    \right],
    \end{equation}
where we used the following asymptotic forms of Bessel functions
    \begin{eqnarray}
    & \dfrac x z \left(\xi^2 +\chi^2\right) \to \dfrac{\pi z}{2}\left(J_1^2(z)+Y_1^2(z)\right), &\\
    & \dfrac{\xi\xi_{\hat \eta}+\chi\chi_{\hat \eta}}{x} \to -\dfrac{\pi z}{4}\left(J_0(z) J_1(z)+Y_0(z) Y_1(z)\right).&
    \end{eqnarray}

For convenience, the spectrum of the occupation number is plotted in Fig.~\ref{fig:black hole}.
In a finite time, the radiation is nonthermal
since the spectrum is quite different from the
Planckian distribution.
When the time goes to infinity,
the spectrum approaches the Planckian distribution.
In particular, in a low frequency region,  Eq.~\eqref{eq:occ num BH infinite} is written as
    \begin{equation}
    \label{eq:Hawking temp}
    N_\infty (\Omega_f)\approx \frac{f'_{\rm out}(R_H)}{4\pi \Omega_f}
    =\frac{\sqrt{M_{\rm out}}}{2\pi\ell~\Omega_f}.
    \end{equation}
Hence, the temperature of the quantum radiation can be read off
from Eq.~\eqref{eq:Hawking temp} by comparing it to
the Planckian distribution in a low frequency as
\begin{equation}
\label{eq:temperature result}
T=\frac{\sqrt{M_{\rm out}}}{2\pi\ell}.
\end{equation}
For $M_{\rm out}> 0$ and $M_{\rm in}=0$, the present analytic account
is compatible with the numerical result~\cite{Saini:2017mur}
that the temperature of the emitted radiation
is very close to the Hawking temperature
as the shell
approaches its own horizon.
Of course, Eq.~\eqref{eq:temperature result} is also valid for $M_{\rm out}> 0$ and $M_{\rm in}=-1$.

\section{conclusion and discussion}
\label{sec:conclusion}
In conclusion, we considered the collapsing dust shell where
the exterior and interior regions of the shell were
described by the AdS geometry
characterized by the different exterior and interior mass parameters of $M_{\rm out}$ and $M_{\rm in}$,
and studied the quantum radiation from the collapsing shell
by using the functional Schr\"odinger formalism.
In the formation of the BTZ black hole where $M_{\rm out} >0$ and $M_{\rm in}=0,-1$,
the wave function in the incipient BTZ black hole limit was fortunately expressed by the analytic form and
so the occupation number of the excited states was calculated exactly.
The spectrum of the occupation number is nonthermal for a finite time, but it is getting close to
the thermal Hawking radiation as time goes to infinity.

Until now, we have studied the collapse of the BTZ black hole for which $M_{\rm out} >0$.
Let us speculate what happens for the collapse where $M_{\rm out} <0$.
As a matter of fact, for $-1< M_{\rm out}< 0$,
the end points of collapse depend on the initial velocity and position, and
can lead to a naked
singularity \cite{Peleg:1994wx,Crisostomo:2003xz,Mann:2006yu}.
To evade the naked singularity attributed to the exterior geometry,
we can instead consider the collapse of the global AdS space as a thermal soliton of $M_{\rm out} =-1$ without the horizon.
Firstly, assuming that the surface energy density $\sigma$ of the shell is positive
in the formation of the global AdS space, then the interior mass parameter should be chosen as $M_{\rm in} <-1$
from Eq.~\eqref{eq:mass rel} so that the interior geometry leads necessarily to the naked singularity.
Of course, if the surface energy density is negative, then
the interior mass parameter can be $M_{\rm in}=0$ for a particular value of $\mu$
where $\mu$ is the constant of motion defined by $\mu =2 \pi R \sigma$; however,
the negative energy density would be unnatural in the classical collapse.
Secondly, the spectrum of the occupation number will appear to be nonthermal
at very late times
when the radius of the shell shrinks to $R \to 0$.
This fact might be expected from the previous work for the collapse of a regular black hole in the absence of the horizon \cite{Um:2019qgc}.
Eventually, the occupation number would not be well-defined when encountering the naked singularity,
which deserves further study.

\acknowledgments
This research was supported by Basic Science Research Program through the National Research Foundation of Korea(NRF) funded by
the
Korea government(MSIP) (NRF-2017R1A2B2006159) and
the Ministry of Education through the Center for Quantum Spacetime (CQUeST) of Sogang University (NRF-2020R1A6A1A03047877).

%%%%%%%%%%%%%%%%%%%%%%%%%%%%%%%%%%%%%%%%%%%%%%%%
%%%%%%%%%%%%%%%             References         %%%%%%%%%%%%%%%%

%\bibliographystyle{mybib}
%\bibliographystyle{apsrev4-1} % PRD
\bibliographystyle{JHEP}       %% JHEP.bst

\bibliography{references}

\end{document}